\documentstyle[11pt,emulateapj]{article}
\lefthead{CHARTAS G.}
\righthead{X-RAY OBSERVATIONS OF GL QUASARS}

\tighten
 
\begin{document}

 

\submitted{The Astrophysical Journal, {\it in press}}
\title{X-ray Observations of Gravitationally Lensed Quasars; Evidence for 
a Hidden Quasar Population}

\author{G. Chartas\altaffilmark{1}}

\altaffiltext{1}{Astronomy and Astrophysics Department, The Pennsylvania
State University, University Park, PA 16802.}

\begin{abstract}

X-ray observations are presented of gravitationally lensed quasars
with redshifts  ranging between 1 and 4. The large magnification factors 
of gravitationally lensed (GL) systems allow us to investigate the properties 
of quasars with X-ray luminosities that are substantially lower than 
those of unlensed ones and also provide an independent means of estimating 
the contribution of faint quasars to the hard X-ray component of the 
cosmic X-ray background. 
Spectral indices have been estimated in the
rest frame energy bands 0.5 - 1keV (soft), 1 - 4keV(mid) and
4 - 20keV(hard). Our spectral analysis indicate a flattening
of the spectral index in the hard band for 2 radio-loud quasars
in the GL quasar sample for which the data have moderate signal-to-noise ratio. 
These results are consistent with the reported spectral properties of non-lensed 
radio-loud quasars, however, there are no indications of spectral hardenning 
towards fainter X-ray fluxes.  

We have identified a large fraction of Broad Absorption Line (BAL)
quasars amongst the GL quasar population. We find that approximately
35$\%$ of radio-quiet GL quasars contain BAL features which is significantly
larger than the 10$\%$ fraction of BAL quasars presently found in
optically selected flux limited quasar samples. We present a simple
model that estimates the effects of attenuation and lens magnification
on the luminosity function of quasars and that explains the 
observed fraction of GL BAL quasars.
These observations suggest that a large fraction of BAL
quasars are missed from flux limited optical surveys.


Modeling of several X-ray observations of the GL BAL quasar PG1115+080
suggests that the observed large X-ray variability may be caused in part by a
variable intrinsic absorber consistent with previously observed variability 
of the BAL troughs in the UV band. The observed large X-ray flux variations 
in PG1115+080 offer the prospect of considerably reducing errors 
in determining the time delay with future X-ray monitoring of this system
and hence constraining the Hubble constant H$_{0}$.



\end{abstract}
 

\section{INTRODUCTION}

Several attempts have been made to characterize the properties of distant
and faint quasars and compare them to those of relatively nearby and bright ones.
(Bechtold et al. 1994; Elvis et al. 1994; Vikhlinin et al. 1995; Cappi et al. 1997; 
Schartel et al. 1996; Laor et al. 1997; Yuan et al. 1998; Brinkmann et al. 1997;
Reeves et al. 1997; Fiore et al. 1998).
The evolution of quasar properties is in part studied by identifying changes in 
spectral properties with redshift.
Information obtained from estimating the X-ray properties of quasars as a 
function of X-ray luminosity and redshift may be useful in constraining 
physical accretion disk models that explain the observed AGN continuum emission.  
The study of X-ray properties of quasars
with relatively low luminosity may also provide clues to the nature of the 
remaining unresolved portion of the hard component of 
the cosmic X-ray background (XRB) (see, for example, Inoue et al. 1996; Ueda 1996; Ueda et al. 1998).

The Gravitational Lensing (GL) effect has been widely employed 
as an analysis tool in a variety of astrophysical situations.
The study of GL systems (GLS) in the radio and optical community has 
proven to be extremely rewarding by providing constraints on cosmological
parameters $\Lambda$ and H$_{0}$ (Kochanek 1996, Falco et al. 1997), by probing the 
evolution of mass-to-light ratios of
high redshift galaxies (Keeton et al. 1998),
by probing the evolution of the interstellar medium of distant 
galaxies (Nadeau et al. 1991), and by determining the total mass, 
spatial extent and ionization conditions of intervening
absorption systems. The study of GL quasars in the X-ray band
has been limited until now, the main limiting factor being
the collecting area and spectral resolution of current X-ray telescopes
combined with the small angular separations of lensed quasar images.
One of the main objectives of this paper is to 
make use of the magnification effect of GL systems  to investigate the X-ray 
properties of faint radio-loud and radio-quiet quasars.

In many cases the available X-ray spectra of distant quasars have relatively low 
signal-to-noise ratio (S/N). This has lead to the development of various techniques
to aid the study of faint quasars
with 0.2 - 2 keV X-ray fluxes below 1 $\times$ 10$^{-14}$ erg s$^{-1}$ cm$^{-2}$.
Most of these observational and analysis techniques  
employed to date to study the evolution and spectral emission 
mechanism of faint quasars are in general a slight variation of
two distinct approaches.
On the one end we find techniques that are
based on summing the individual spectra of many faint X-ray sources 
taken from a large and complete sample 
(see for example Schartel et al. 1996; Vikhlinin et al. 1995). 
The goal of stacking is to obtain a single, high signal-to-noise ratio
spectrum that contains enough counts 
to allow spectral fitting with quasar emission models.
In some cases, where the initial sample is large enough, the 
spectra can be summed into bins of X-ray flux, redshift 
and radio luminosity class.
Schartel et al. 1996 applied the stacking technique to a complete 
quasar sample and found that the mean spectral index for the 
stacked radio-quiet quasar spectrum was significantly ($\sim2\sigma$) steeper  
than that of the stacked radio-loud quasar spectrum. 
Several of the assumptions made in this analysis which were related to the 
general properties of the quasars may, however, strongly influence 
the further interpretation of the results.
It was assumed, for example, that quasar spectra follow single power-laws 
over rest frame energies ranging between (1+z)$E_{min}$ and (1+z)$E_{max}$, 
where $E_{min}$ and $E_{max}$ are the minimum and maximum energy bounds, 
respectively, of the bandpass of the X-ray observatory and z is the redshift of the quasar.
Analyzed in the observer's frame, quasar spectra with concave spectral slopes of increasing 
redshift would appear to be flatter as a consequence of the cosmological redshift.
Any implications of evolutionary change within the quasar spectrum 
derived from the analysis of stacked spectra in observed frames 
would need to include the effect of cosmological redshift.
Detailed spectra of quasars however indicate in general 
the presence of several components each associated to 
a different physical process. For example, several known processes contributing to 
the observed X-ray spectra are Compton reflection of photons in the
disk coronae by the disk which becomes significant
at rest energies above $\sim$ 10keV, inverse Compton scattering of photons in the 
disk coronae by UV photons originating from the disk resulting in a boost of photon 
energies from the UV range into the soft X-ray range (this is the mechanism that produces 
the observed power-law spectrum in the 2-10keV range),  
accretion disk emission, absorption by highly ionized gas
(warm absorbers), beamed X-ray emission from jets which may be a large contributor for distant radio-loud quasars, 
absorption by accretion disk winds, and intervening
absorption by damped Lyman alpha systems. 
The stacked spectrum therefore contains contributions from quasars 
of different redshifts and possibly spectral shapes that
make the interpretation of the results difficult. 

Vikhlinin et al. 1995, using the ROSAT EMSS sample of 2678 sources, produced 
stacked spectra within several flux bins. They find a significant continuous 
flattening of the fitted spectral slopes from higher towards lower X-ray fluxes.  
One interesting result of their study is that the spectral 
slope at the very faint end is approximately equal to the slope 
of the hard (2 - 10keV) X-ray background. The unknown nature of many of the 
point sources included in the Vikhlinin sample,
the inclusion of sources with different redshifts and the 
calculation of observed rest spectral indices complicates the interpretation
of the results.

A second technique used to study the general properties of quasars
is based on obtaining deep X-ray observations of a few quasars. One advantage
of such an approach is that the properties 
of individual quasars are not smeared out as with stacking methods. The faint fluxes
however require extremely long observing times to achieve useable S/N.
When total counts are low the quasar X-ray spectra are commonly  characterized by a hardness ratio
defined as R = {(H - S)}/{(H + S)}, where H and S are the number of counts within
some defined hard and soft energy band in the observer's frame respectively.

In this paper we outline an alternative approach to investigating the
emission mechanism of radio-loud and quiet quasars at high redshift.
The gravitational lensing magnification of distant quasars
allows us to investigate the X-ray properties
of quasars with luminosities relatively lower than those of unlensed quasars
of similar redshifts. The amplification factors produced by  
lensing depends on the geometry of the lensing system and for
our sample range between 2 and 30. The moderate-S/N spectra of our sample 
allow us to employ spectral models with multiple power-law slopes and perform fits in rest-frame
energy bands. Our analysis makes use of the GL amplification effect
to extend the study of quasar properties to unlensed X-ray flux levels
as low as a few $\times$ 10$^{-16}$ erg s$^{-1}$ cm$^{-2}$. The limiting sensitivity of
the ROSAT All-Sky Survey, for example, on which many recent studies are based
is a few $\times$ 10$^{-13}$ erg s$^{-1}$ cm$^{-2}$.

For a GL system with a lens that can be modeled well with
a singular isothermal sphere (SIS) model the amplification 
is straightforward to derive analytically. In most observed cases,
however, the deflector is a galaxy or cluster of galaxies with 
a gravitational potential that does not follow the SIS model 
and more sophisticated potential models need to be invoked to 
successfully model these GL systems.
To estimate the intrinsic X-ray luminosity of
GL quasars in our sample we have incorporated magnification factors 
determined from modeling of the GL systems with a variety of
lens potentials. 

We performed fits of spectral models to the X-ray data in three
rest energy bins, soft from 0.5 - 1 keV, mid from 1 - 4 keV,
and high from 4 - 20 keV. Working in the quasar rest frame as opposed to
the observers rest frame allows us to distinguish between 
true spectral evolution of quasars
with redshift and apparent change in quasar spectra  
due to the cosmological redshift of quasar spectra through
a fixed energy window in the observer's rest frame. 

Our search of the ROSAT and ASCA archives yielded 16
GL systems detected in X-rays out of a total of approximately 40 GL candidates.
Six of the GL quasars have observed X-ray spectra of medium-S/N.
Our search for X-ray counterparts to known GL quasars resulted in the identification of the 
relatively X-ray bright radio-quiet quasar SBS0909+532 with an estimated 
0.2-2 keV flux of about 7 $\times$ 10$^{-13}$ erg s$^{-1}$ cm$^{-2}$.
Another interesting result of our search was the identification
of a relatively large fraction of radio-quiet GL quasars with 
BAL quasars. In particular we find that at least 35\%
of the known radio-quiet GL quasars are BAL quasars. This value
is significantly larger than the $\sim$ 10\% value presently
quoted from optical surveys.

In section 2 we present details of X-ray observations
of GL quasars and describe the analysis techniques used to 
extract and fit the X-ray spectra. Estimates of the flux magnification factors
and unlensed luminosities for the GL systems studied in this paper are 
presented in section 3. A description of the properties of 
each GL quasar is presented in section 4. Included in this section are
results from spectral modeling of several X-ray observations of the variable GL BAL quasar
PG1115+080.  Finally, in section 5
we summarize the spectral properties of faint quasars as implied by
spectral fits to a sample of GL quasar spectra
and provide a plausible explanation for the
apparently large fraction of GL BAL quasars that we observe.

\section{X-RAY OBSERVATIONS AND DATA ANALYSIS}
The X-ray observations presented here were performed with the ROSAT
and ASCA observatories. Results for the spectral analyses in the
X-ray band for the GL quasars
Q0957+561, HE1104-1805,  PKS1830-211 and Q1413+117 have already been published 
(Chartas et al. 1995, 1998; Reimers et al. 1995; Mathur et al. 1997; 
Green \& Mathur, 1996) 
while results from X-ray spectral analyses for the quasars SBS0909+532,
B1422+231, PG1115+080, 1208+1011 and QJ0240-343 
are presented here for the first time. We have included in Table 1 
several additional GL systems observed in the X-ray band. These observations
however yielded either very low-S/N detections or were 
made with the ROSAT HRI which provides very limited spectral information.
X-ray spectra of the GL quasars Q0957+561, 1422+231, HE1104-1805,
SBS0909+532 and PG1115+080 with best fit models are presented in Figure 1
through Figure 5 respectively.  

\begin{figure*}[t]
\plotfiddle{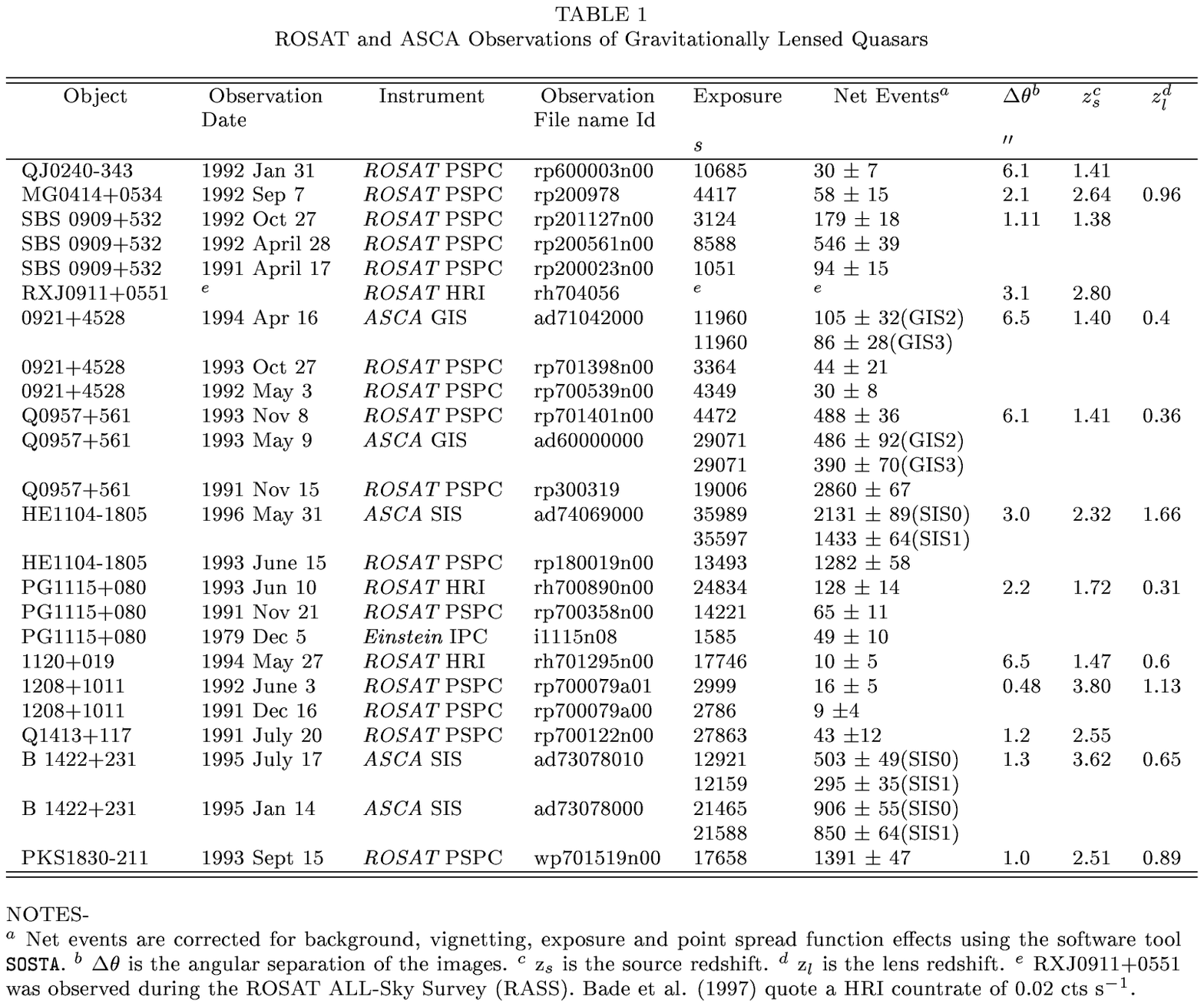}{5.5in}{0}{100.}{100.}{-340}{-250}
\end{figure*}

\begin{figure*}[t]
\plotfiddle{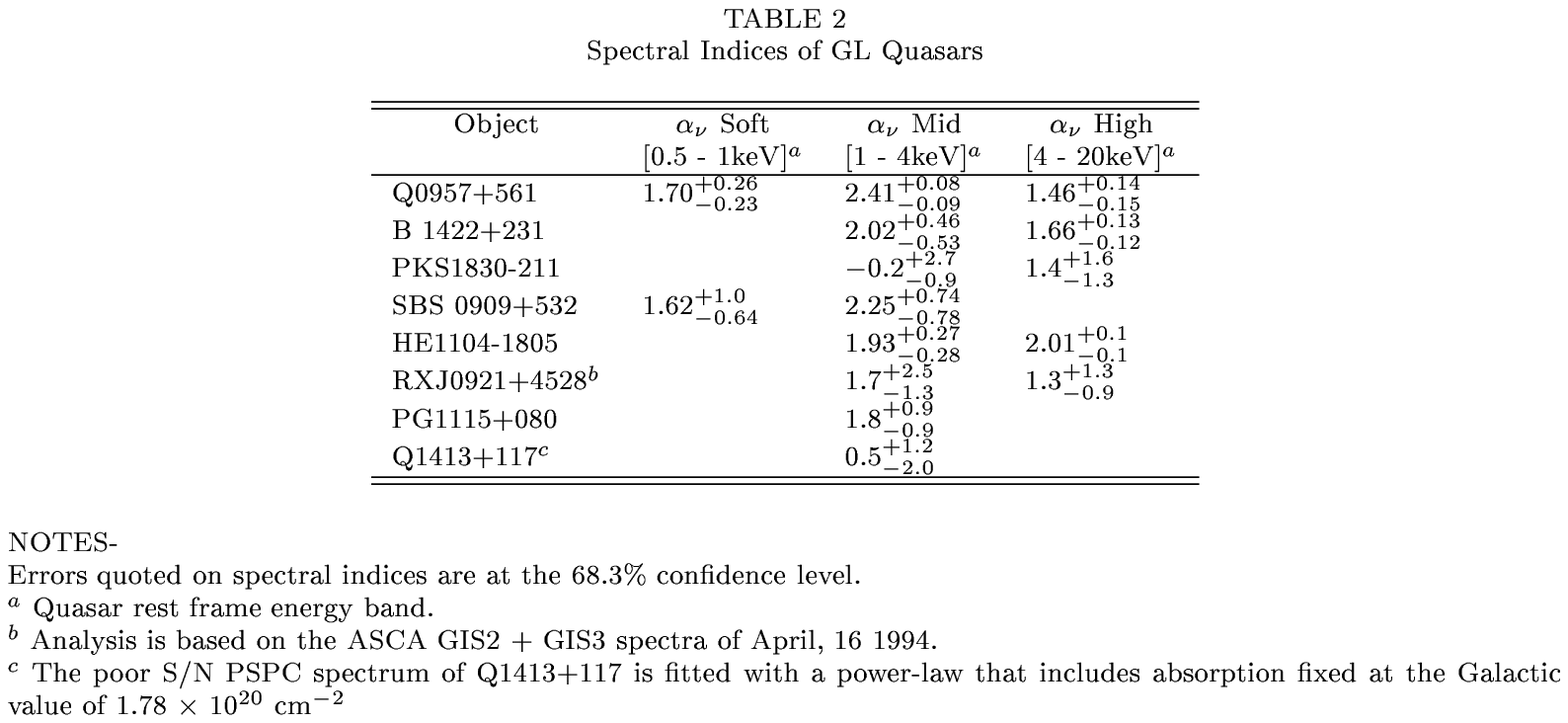}{3.4in}{0}{100.}{100.}{-340}{-430}
\end{figure*}

\begin{figure*}[t]
\plotfiddle{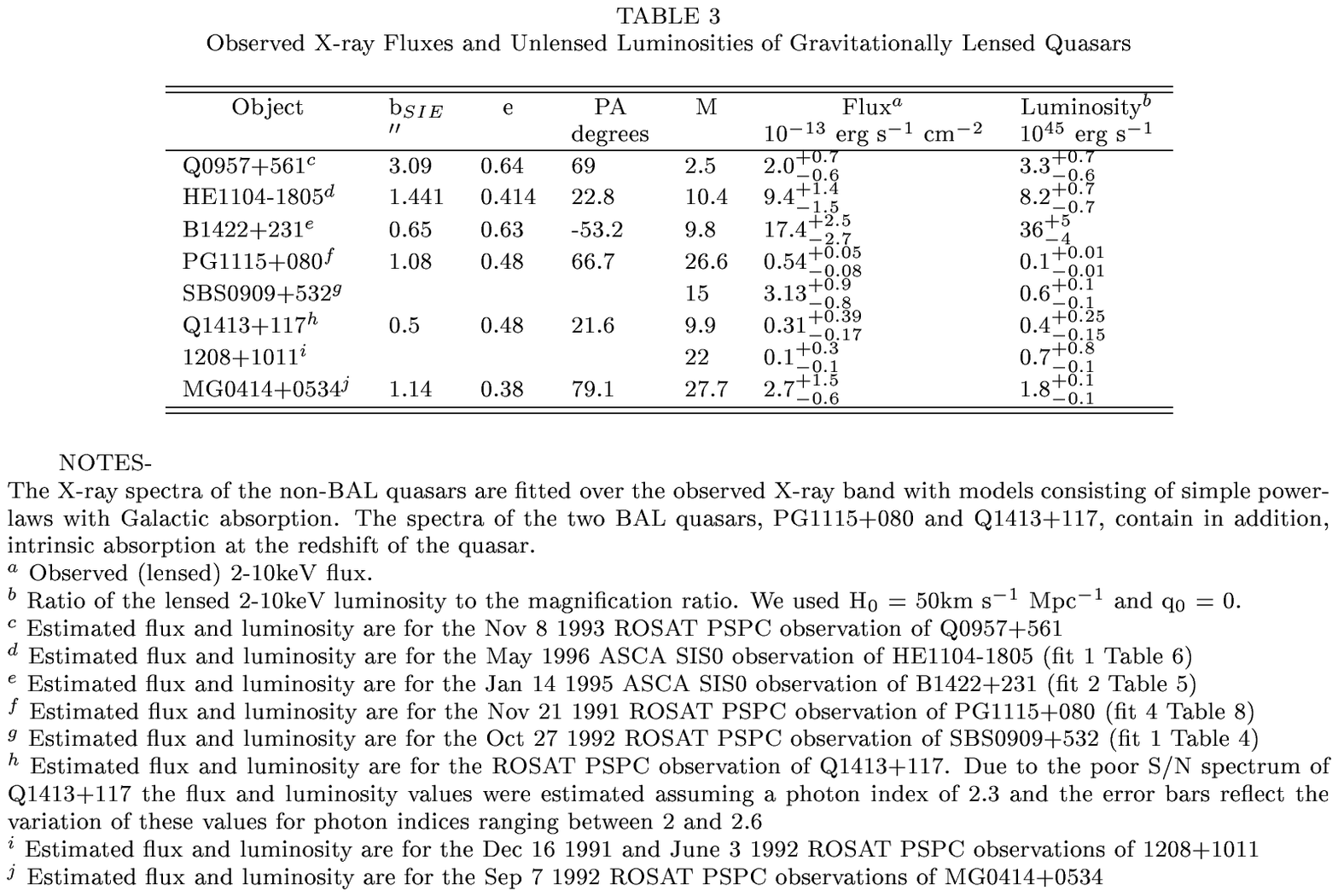}{4.3in}{0}{100.}{100.}{-340}{-345}
\end{figure*}

\begin{figure*}[t]
\plotfiddle{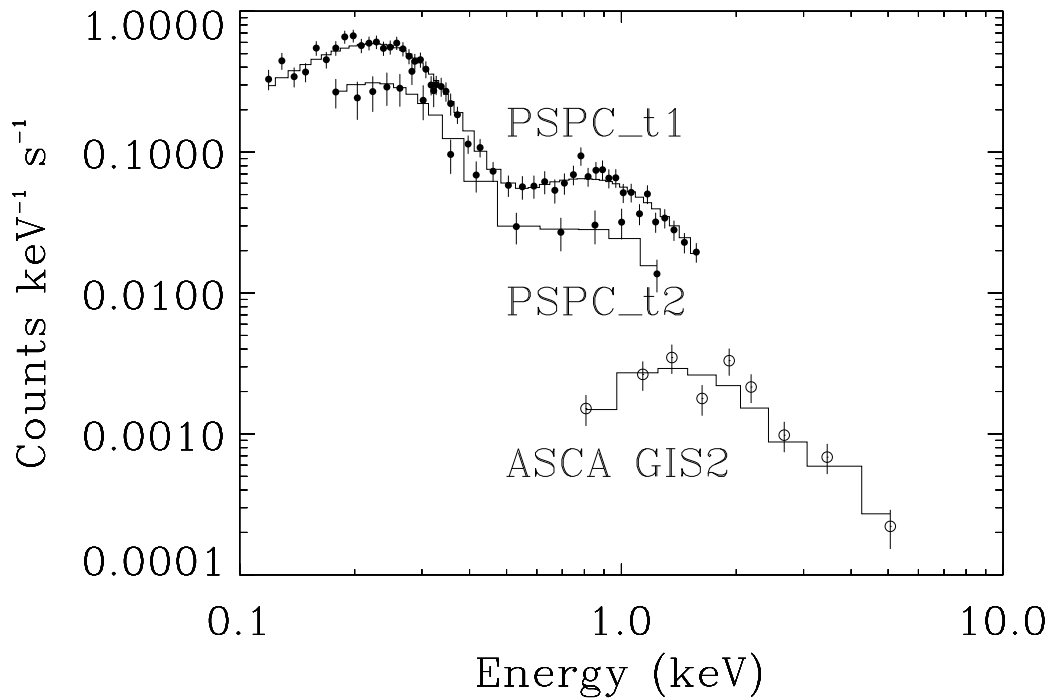}{4.19in}{90}{60.}{60.}{230}{-20}
\protect\caption
{\footnotesize
ROSAT PSPC and ASCA GIS spectra of Q0957+561 with best fit models.
 \label{fig:fig1} 
}
\end{figure*}

\begin{figure*}[t]
\plotfiddle{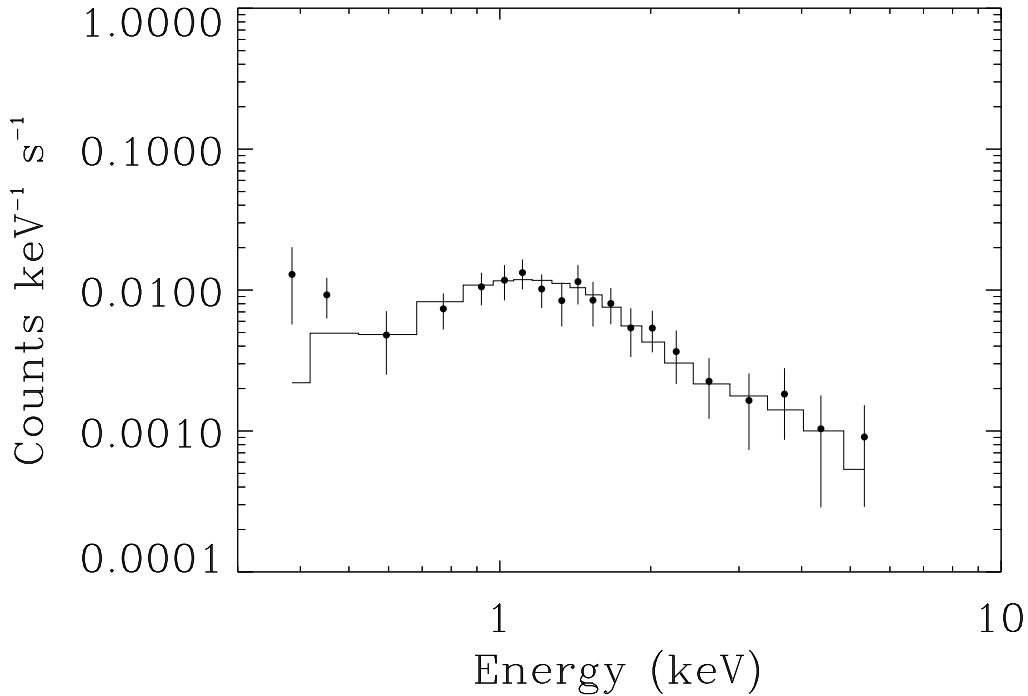}{4.185in}{90}{60.}{60.}{230}{-20}
\protect\caption
{\footnotesize
ASCA SIS spectrum of 1422+231 with best fit models.
 \label{fig:fig2} 
}
\end{figure*}

\begin{figure*}[t]
\plotfiddle{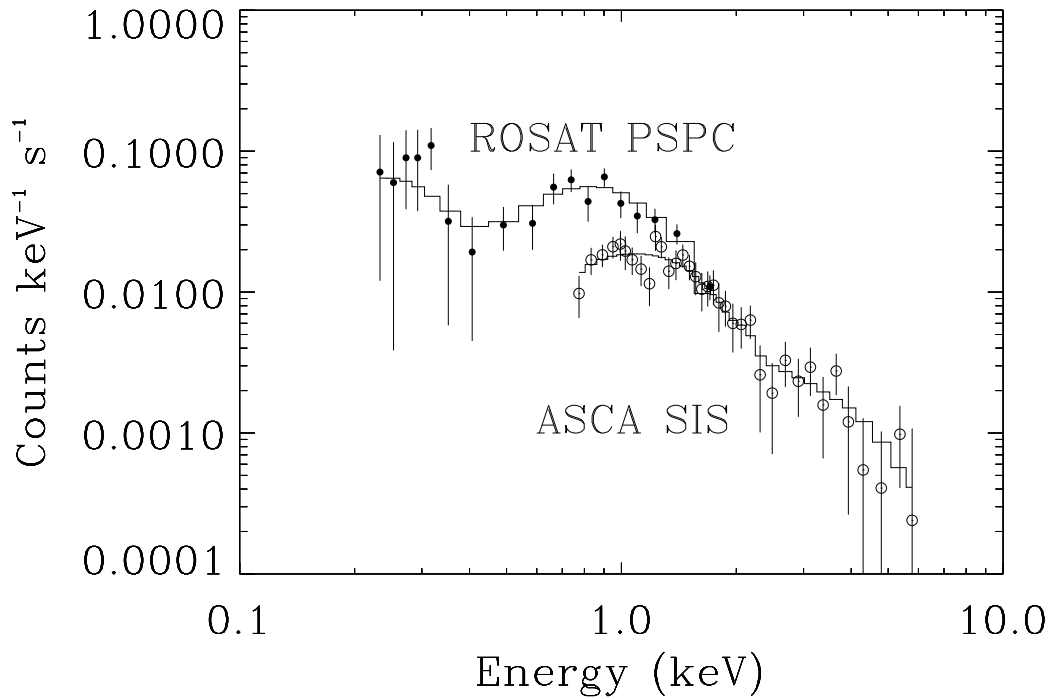}{4.185in}{90}{60.}{60.}{230}{-20}
\protect\caption
{\footnotesize
ROSAT PSPC and ASCA SIS spectra of HE1104-1805 with best fit models.
 \label{fig:fig3} 
}
\end{figure*}

\begin{figure*}[t]
\plotfiddle{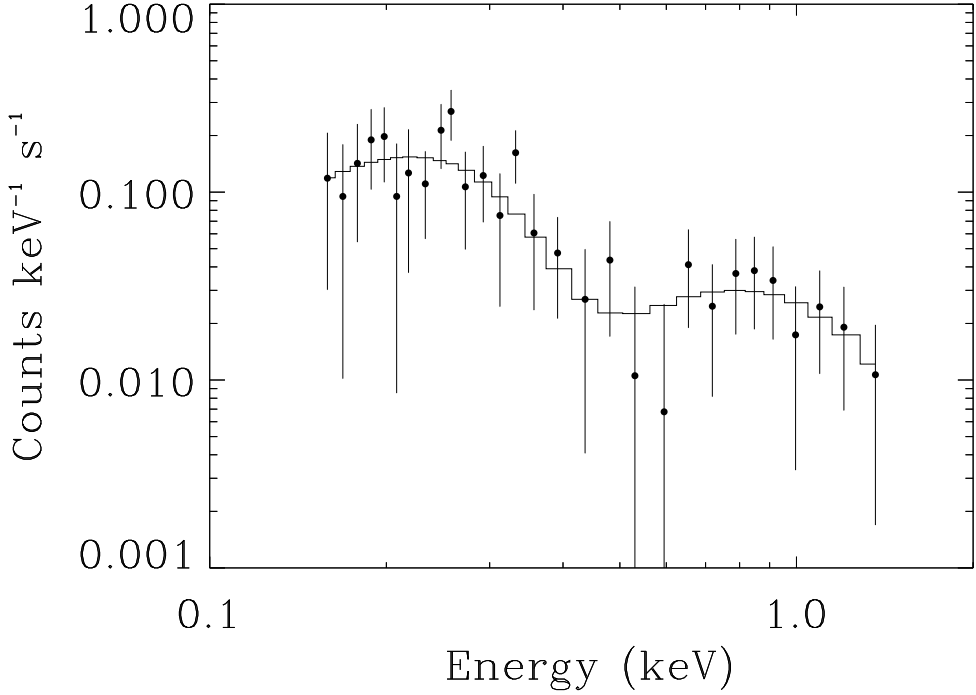}{4.185in}{90}{60.}{60.}{230}{-20}
\protect\caption
{\footnotesize
ROSAT PSPC spectrum of SBS0909+532 with best fit model.
 \label{fig:fig4} 
}
\end{figure*}

\begin{figure*}[t]
\plotfiddle{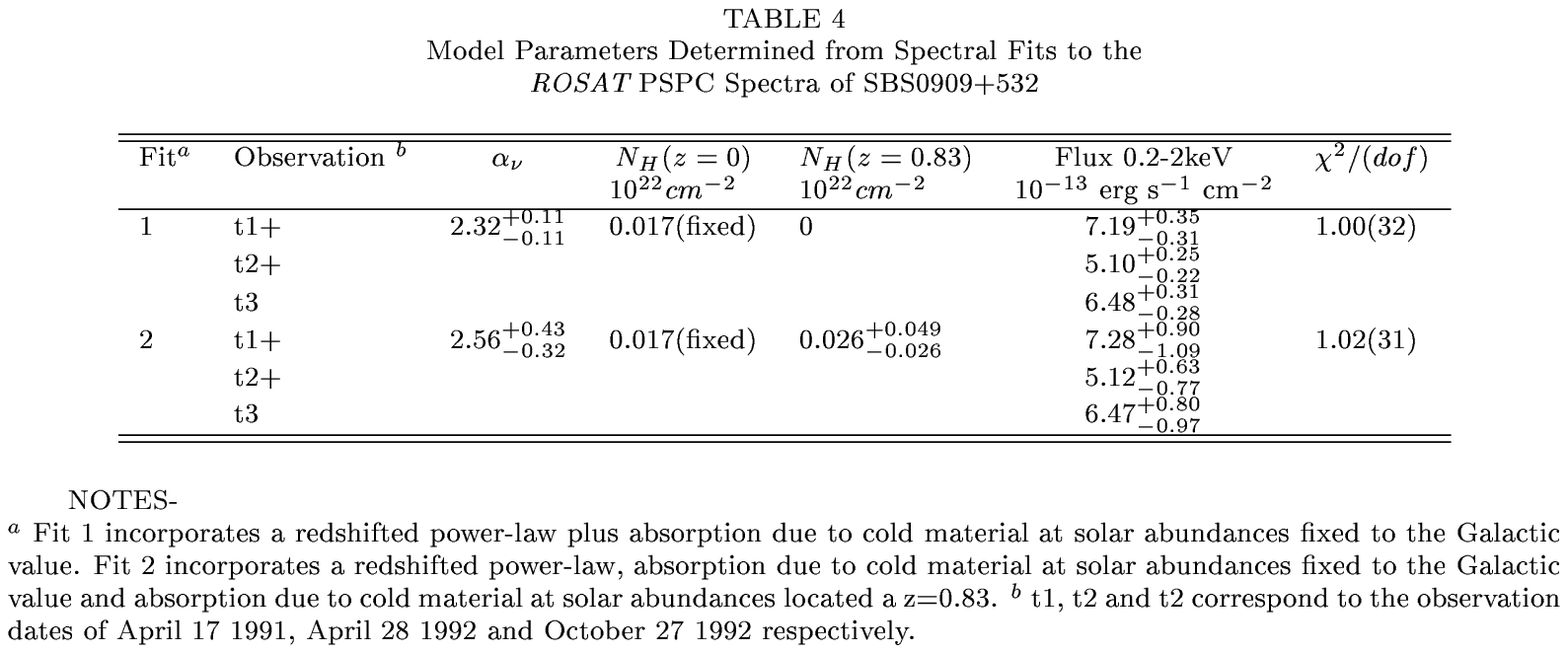}{3.in}{0}{100.}{100.}{-340}{-450}
\end{figure*}

\begin{figure*}[t]
\plotfiddle{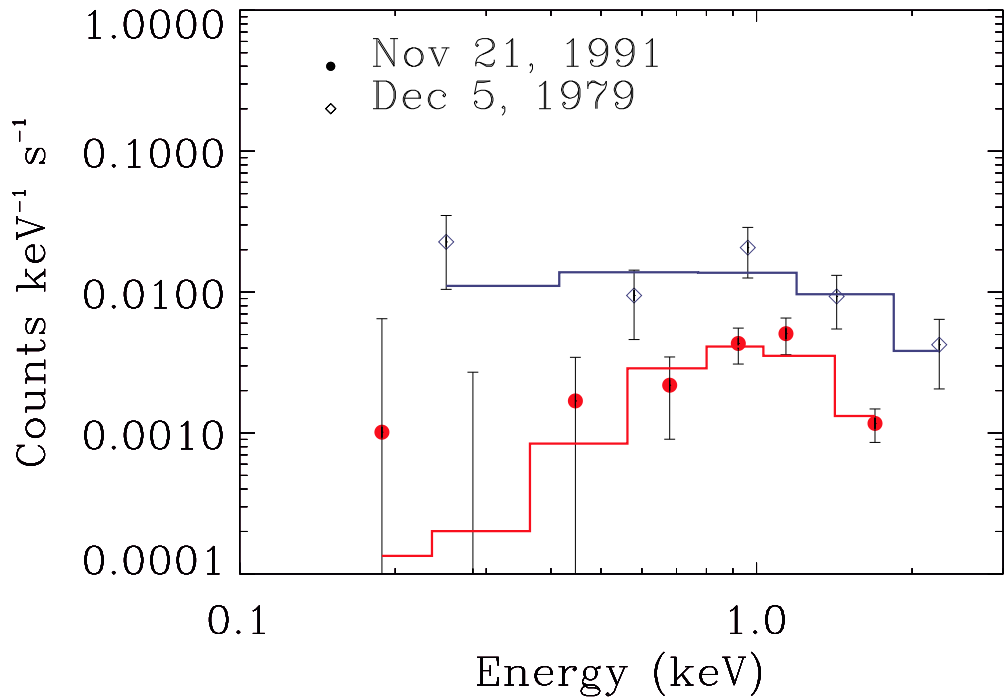}{4.185in}{90}{60.}{60.}{230}{-20}
\protect\caption
{\footnotesize
Einstein IPC and ROSAT PSPC
spectra of PG1115+080 observed on Dec 5, 1979 and Nov 21, 1991 respectively, accompanied by best fit models.
 \label{fig:fig5} 
}
\end{figure*}

\begin{figure*}[t]
\plotfiddle{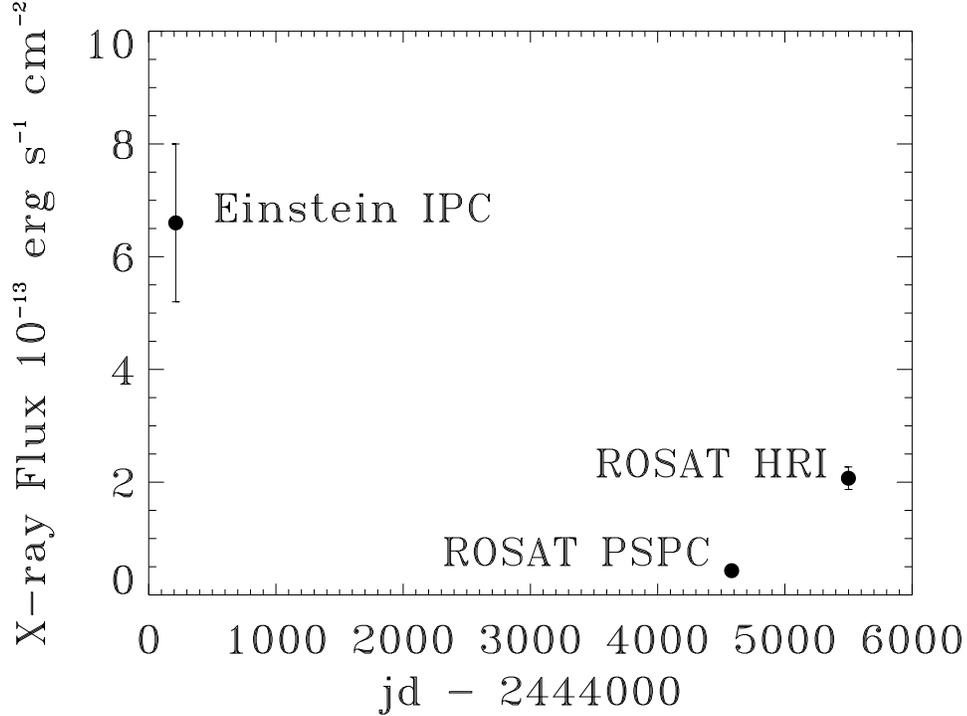}{4.185in}{90}{60.}{60.}{230}{-20}
\protect\caption
{\footnotesize
Estimated 0.2-2keV flux levels 
of PG1115+080 for the three available X-ray observations.
 \label{fig:fig6} 
}
\end{figure*}

\begin{figure*}[t]
\plotfiddle{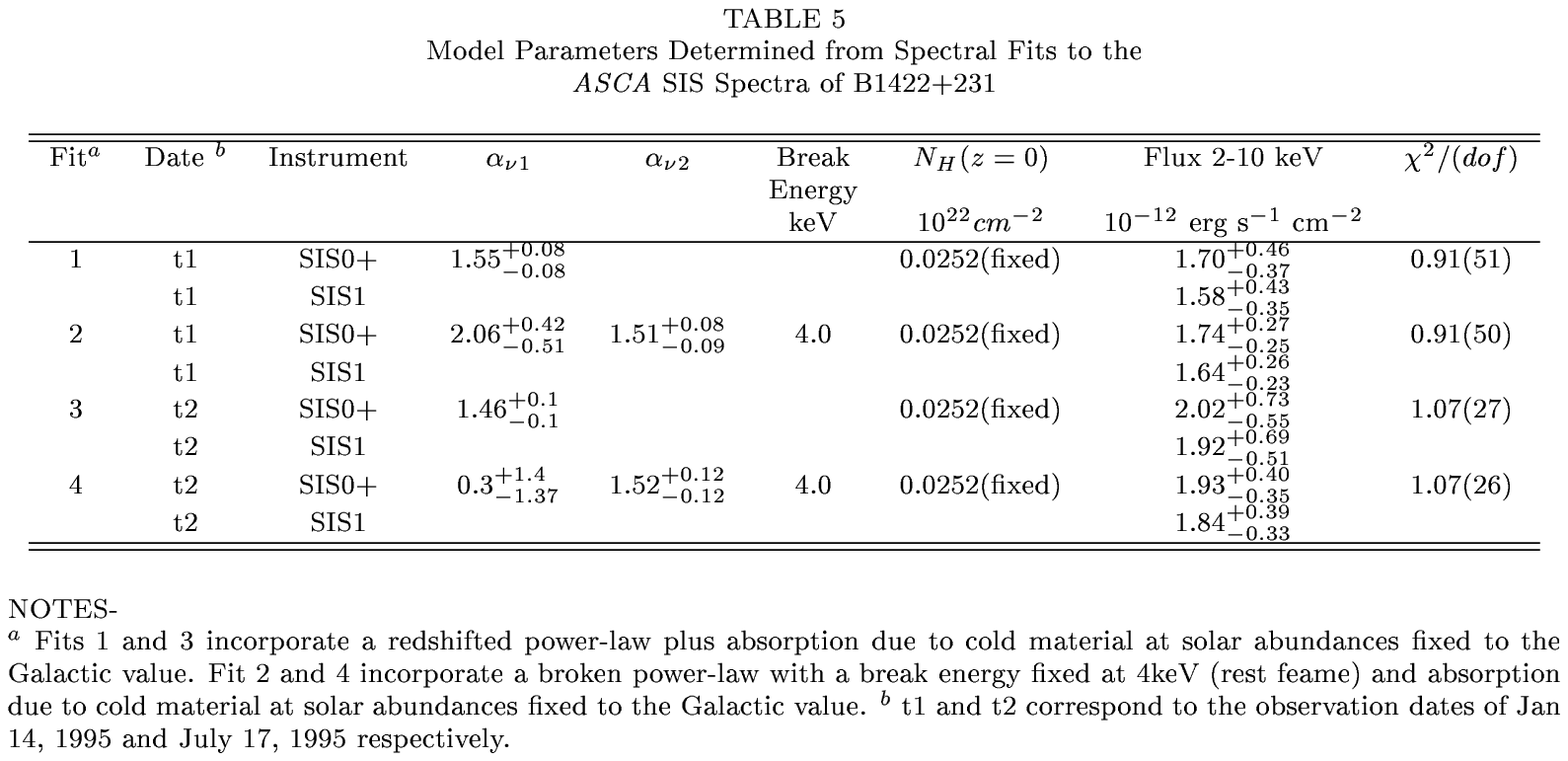}{3.in}{0}{100.}{100.}{-340}{-440}
\end{figure*}

\begin{figure*}[t]
\plotfiddle{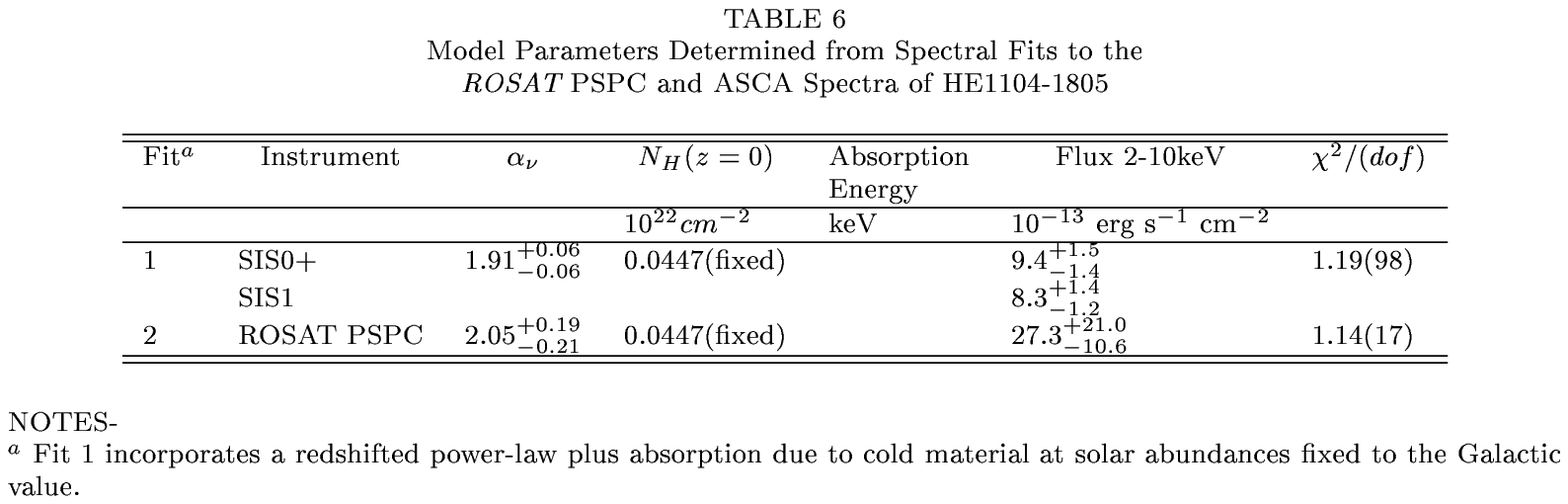}{2.3in}{0}{100.}{100.}{-340}{-495}
\end{figure*}

\begin{figure*}[t]
\plotfiddle{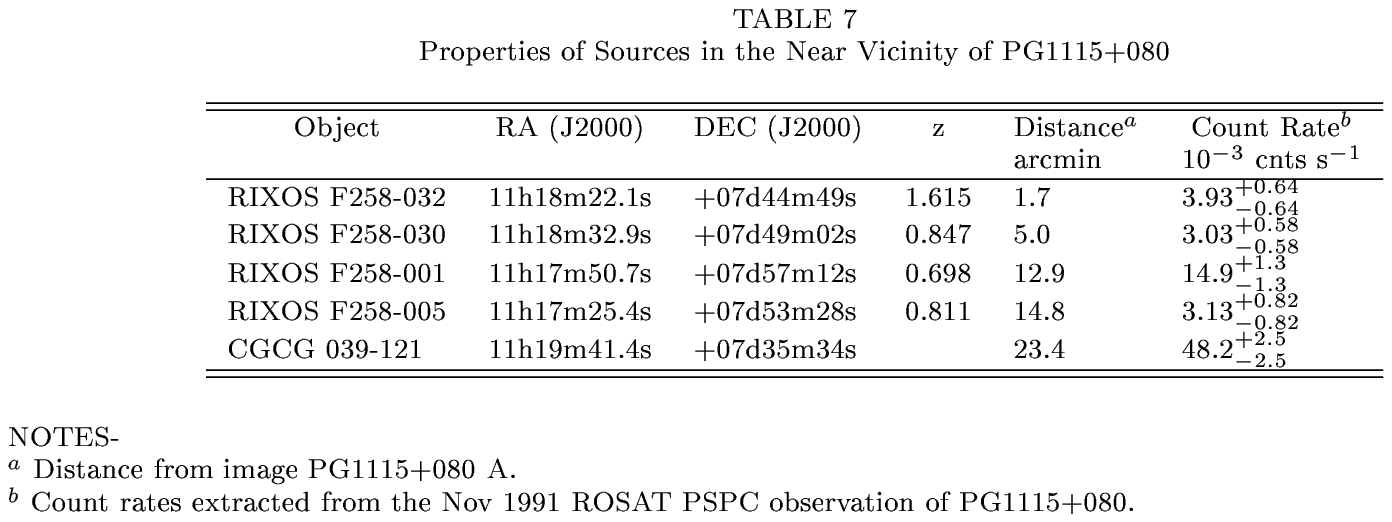}{2.2in}{0}{100.}{100.}{-340}{-490}
\end{figure*}

\begin{figure*}[t]
\plotfiddle{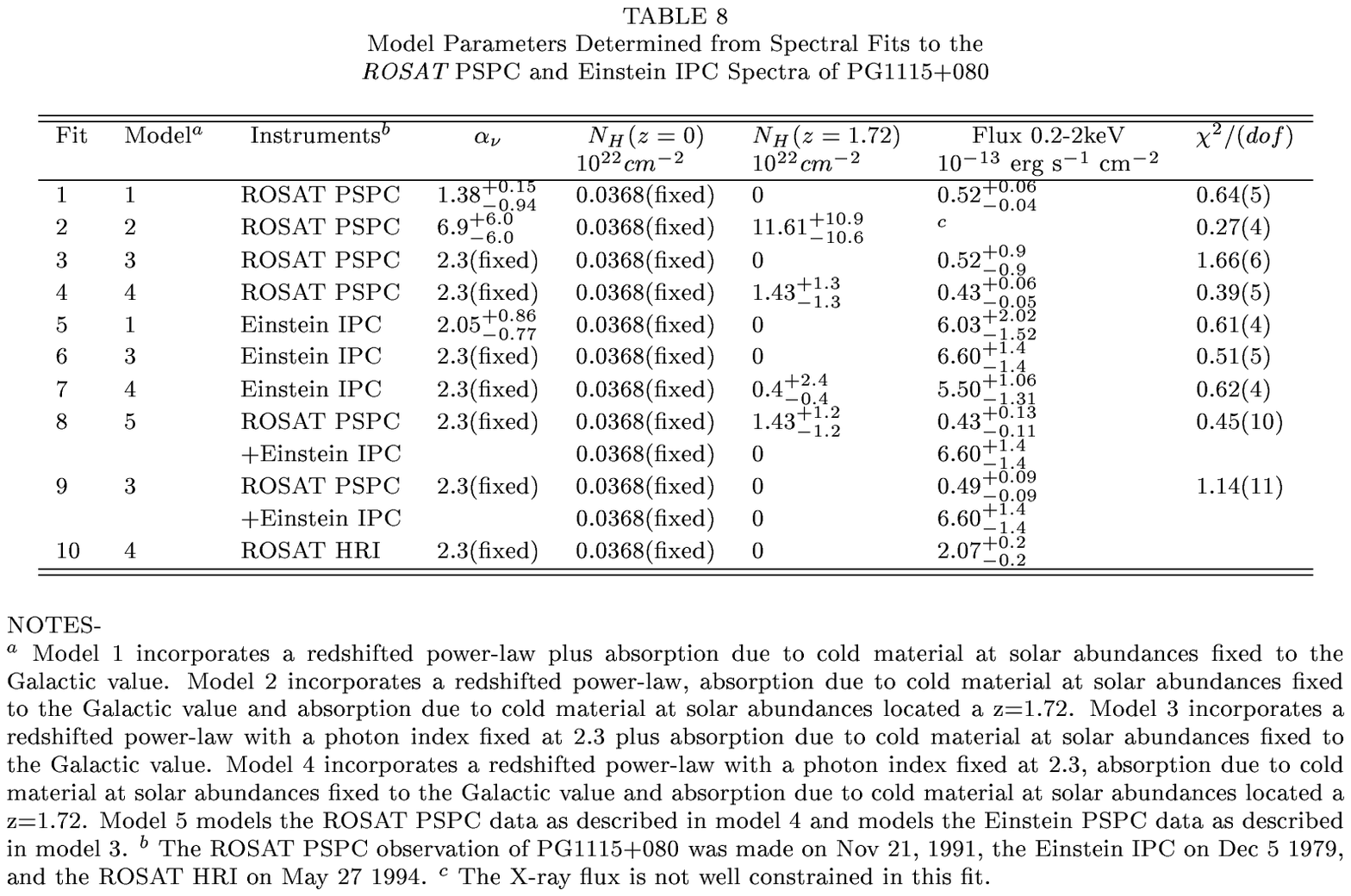}{4.in}{0}{100.}{100.}{-340}{-380}
\end{figure*}

\begin{figure*}[t]
\plotfiddle{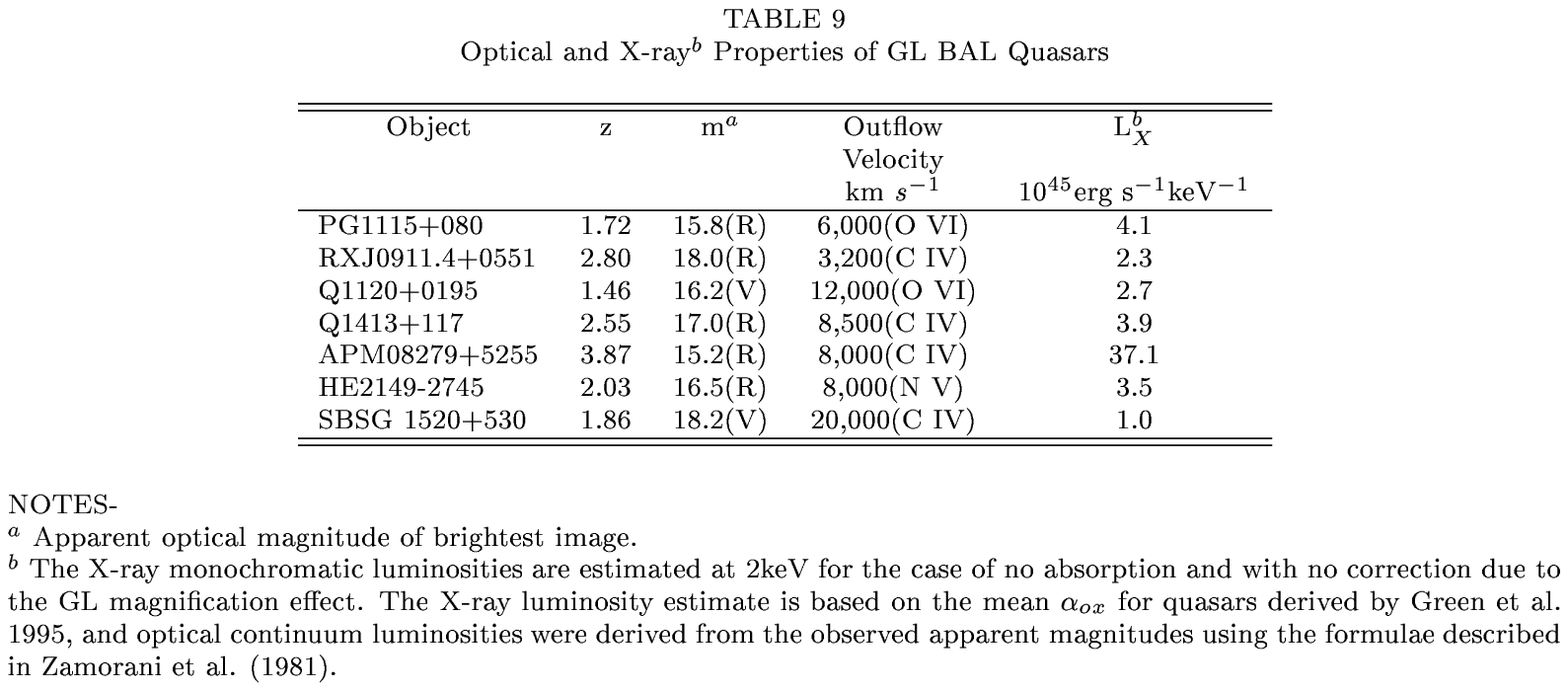}{3.5in}{0}{100.}{100.}{-340}{-440}
\end{figure*}

The wide separation quasar pairs MG 0023+171, Q1120+0195, 
LBQS 1429-008 and QJ0240-343 are considered problematic GL candidates 
that may be binary quasars and not gravitational lenses (Kochanek et al. 1999).  
However, recent STIS spectroscopy of Q1120+0195(UM425) has revealed
broad absorption line features in both lensed images thus confirming the lens 
nature of this wide separation system (Smette et al. 1998).
The origin of the detected X-ray emission for the GL systems RXJ0921+4528 and RXJ0911.4+0551
cannot be determined from the presently available poor S/N  ASCA GIS and ROSAT HRI observations
respectively. The most likely origins are either the lensed quasar and/or a possible lensing cluster.
In Table 1 we list the ASCA and ROSAT observations of GL quasars with detected X-rays.

The spatial resolution for on axis pointing of the ROSAT HRI, 
ROSAT PSPC and ASCA GIS is about 5$''$, 25$''$ and 3$'$ respectively. 
Thus only for the ROSAT HRI observation of Q0957+561 was it 
possible to resolve the lensed X-ray images (Chartas et al. 1995).

For the data reduction of the ASCA and ROSAT observations we used the FTOOLS 
package XSELECT. The ASCA SIS data used in this
study were all taken in BRIGHT mode and high, medium and 
low bit rate data were included in the analysis.
We created response matrices and ancillary files for each chip 
using the FTOOLS tasks {\tt sisrmg} and {\tt ascaarf} respectively.
For the ROSAT PSPC data reduction we used the response matrix
{\tt pspcb\_gain2\_256.rmf} supplied by the ROSAT GOF and created 
ancillary files using the FTOOLS task {\tt pcarf}.
Net events shown in Table 1 are corrected for
background, vignetting, exposure and point spread function effects
using the source analysis tool {\it SOSTA} (part of the software package XIMAGE). 



For the spectral analyses we are mostly limited by the energy resolution and 
counting statistics to fitting simple absorbed power-law models to the data.
In addition to spectral fits in the quasar rest frame 
we also performed spectral fits in the observer's 
reference frame to facilitate the comparison of our results to those of previous 
studies. In Table 2 we show the results from fits of absorbed power-law 
models within three quasar rest frame energy intervals. 
In most cases no data are available in the soft interval
since the corresponding observed energy interval is redshifted below the 
low energy quantum efficiency cutoff of the ROSAT XRT/PSPC.

To estimate any significant difference of the spectral properties of
our GL sample from those of unlensed quasar samples we computed the merit 
function, ${\chi^{2}}\over{N}$, defined by the expression,

\begin{equation}
{{\chi^{2}}\over{N}} = \sum_{i=0}^{N} {{[\alpha_{\nu,GL}(i) - \alpha_{\nu,UL}(i)]^{2}}\over{{\sigma_{UL}}^{2}(i)}}
\end{equation}

\noindent
where $\alpha_{\nu,GL}$(i) and $\alpha_{\nu,UL}$(i) are the spectral indices 
of the GL and unlensed quasar sample respectively, $\sigma_{UL}(i)$ are the 
errors of the spectral indices of the unlensed samples
and N is the number of spectral indices compared. 
We computed the merit function between our data set and that of Fiore et al. 1998.
For the comparison we computed the spectral indices $\alpha_{S}$(0.1 - 0.8 keV) and
$\alpha_{H}$(0.4 - 2.4 keV) in the quasar observed frames. 
Incorporating the 1$\sigma$ uncertainties in the fitted values for the GL spectral
indices we obtain a distribution of values for the merit function with a most likely 
value of ${\chi^{2}}\over{N}$ $\sim$ 0.5 with N = 7. 
We therefore conclude that there is no significant difference 
at the $\Delta\alpha_\nu$ = 0.3 level between our lensed sample 
and the Fiore et al. 1998 sample.      


Spectral modeling of the radio-loud and (non-BAL) radio-quiet quasars
of our GL sample that have unlensed X-ray fluxes
ranging between 3 $\times$ 10$^{-16}$ and
1 $\times$ 10$^{-12}$ erg s$^{-1}$ cm$^{-2}$
have indices that are consistent with those of brighter quasars (see Figure 7). 
In spite of the fact that most of the quasars in our sample have 
intervening absorption systems (see comments on individual systems)
we find that the estimated photon indices for the very faint non-BAL quasars
of our sample do not approach the level of 1.4 of the hard X-ray background.

In Figure 8 we also show that the photon indices of our GL quasar sample
do not show any signs of hardenning over a range of three orders of magnitude
in unlensed 2-10keV luminosity.
The three apparently harder spectra of Figure 7
and Figure 8 correspond to two BAL quasars and one absorbed
blazar of our sample. The X-ray spectra of BAL quasars
that are modeled as power-laws with Galactic absorption and no intrinsic 
absorption will erroneously imply relatively low photon indices. For example,
spectral fits of a simple power-law plus Galactic absorption model to the X-ray spectrum 
of PG1115+080 (Fit 1, Table 8) yields a relatively low photon index of 1.4.
The presently available spectra of the two BAL quasars of our sample
have poor S/N and can not provide significant constrains
on the intrinsic absorber column densities.

\section{QUASAR UNLENSED LUMINOSITY}

The apparent surface brightness of gravitationally lensed images
is a conserved quantity, however the observed X-ray flux
is amplified due to the geometric distortion of the GL images.
The lensed quasar images of our sample are not spatially resolved so 
we only observe the total magnification of the X-ray flux and
not the spatial distortion of the image. Gravitational lensing is
in general an achromatic effect, however possible differential absorption
in the multiple images, microlensing from stars in the lens
galaxy and source spectral variability combined with the expected time delay between 
photon arrival for each image may produce distinct features in the multiply lensed spectrum
of the quasar.

We estimated the unlensed X-ray luminosity of the quasars in our sample 
by scaling the lensed luminosity determined from the spectral fits to
the GL magnification factors. The magnification parameters were derived 
from fits of singular isothermal ellipsoid SIE lens models 
(Keeton, Kochanek, \& Falco, 1997) to
optical and radio observables (e.g. image and lens positions and flux ratios) 
and incorporating the best fit parameters to derive the 
convergence, $\kappa$, of the lens.

For a SIE lens the convergence $\kappa({\bf x})$ is given by,

\begin{equation}
\kappa({\bf x}) = {{b}\over{x\sqrt{1 + e\;\cos{(2(\theta - \theta_{0}))}}}}
\end{equation}
where $b$ is the best-fit critical radius, $x$ is the distance from the lens galaxy center,
$\theta$ is the position angle of point ${\bf x}$ with respect to the lens galaxy,
$e$ is the ellipticity parameter of the lens and $\theta_{0}$ is
the major axis position angle. 

The magnification $\mu(x)$ of each lensed image for an SIE lens is given by 
(Kormann, Schneider, and Bartelmann, 1994),
\begin{equation}
\mu(x) = {1 \over{ (1 - 2\kappa)}}
\end{equation}

In Table 3 we provide GL model parameters and magnification factors 
for several GL systems.

\section{COMMENTS ON INDIVIDUAL SOURCES}

\subsection{The Newly Identified GL X-Ray Source SBS 0909+532}
The radio-quiet quasar SBS0909+532 was recently identified as a candidate
gravitational lens system with a source redshift of 1.377, and an
image separation of 1.107$''$. The lens has not yet been clearly identified,
however, GL statistics place the most likely redshift
for the lens galaxy at z$_{l}$ $\simeq$ 0.5 with 90$\%$ confidence bounds
of 0.18 $<$ z$_{l}$ $<$ 0.83 (Kochanek et al. 1997).
Optical spectroscopy (Kochanek et al. 1997; Oscoz et al. 1997) has identified heavy element
absorption lines of CIII, FeII and Mg II at z = 0.83. 
The optical data at this point cannot clearly discern whether the
heavy-element absorber is associated with the lensing galaxy.

We searched the HEASARC archive and found a bright X-ray source within
7$''$ of the optical location of SBS0909+532, well within the
error bars of the ROSAT pointing accuracy of $\sim$ 30$''$. The position of the 
X-ray counterpart as determined using the {\it detect} routine, which is part of the XIMAGE
software package, is 09h 13m 1.7s, 52$^{\circ}$ 59$'$ 39.5$''$ (J2000), whereas  the optical 
source coordinates of SBS0909+532 are 09h 13m 2.4s, 52$^{\circ}$ 59$'$ 36.4$''$ (J2000).
 
This X-ray counterpart was observed serendipitously with the ROSAT PSPC
on April 17 1991, April 28 1992 and October 27 1992 with detected count rates
of 0.057 $\pm$ 0.006, 0.064 $\pm$ 0.005 and 0.09 $\pm$ 0.01 cnts s$^{-1}$ respectively.


We performed simultaneous spectral fits in the observer frame 
to the three ROSAT PSPC observations. 
The results are summarized in Table 4. We considered two types of spectral models. 
In fit 1 of Table 4 we incorporated a redshifted power-law
plus Galactic absorption and in fit 2 we included additional
absorption at a redshift of 0.83 (possible lens redshift). 
Fits 1 and 2  yield acceptable reduced $\chi^{2}$(dof) of 1.00(32) and 1.02(31) respectively.
We can rule out absorption columns at z = 0.83 of more than 7.5 $\times$ 10$^{20}$ cm$^{-2}$
at the 68.3\% confidence level.

We also performed spectral fits in the quasar rest frame bands
0.5 - 1keV (soft band) and 1 - 4 keV (mid band). 
Simple spectral fits with an absorbed power-law assuming Galactic absorption
of 1.72 $\times$ 10$^{20}$ cm$^{-2}$ result in spectral indices of 
$1.62^{+1.0}_{-0.64}$ and $2.25^{+0.74}_{-0.78}$ for the soft and mid bands 
respectively. All errors quoted in this paper are at the 68.3$\%$ 
confidence level unless mentioned otherwise. No X-ray data are
presently available for the high energy band  4 - 20 keV. The ROSAT PSPC 
can only detect photons with energies up to about 3 keV in 
the rest frame of SBS0909+532.
In Figure 4 we show the ROSAT PSPC spectrum of 
SBS0909+532 together with the best fit absorbed power-law model.

\subsection{B1422+231}
B1422+231 is a well studied quadrupole GLS with the lensed source being a 
radio-loud quasar at a redshift of 3.62 (Patnaik et al. 1992) and the lens consisting of
a group of galaxies at a redshift of about 0.34 (Tonry, 1998).  
CIV doublets were found at redshifts of 3.091, 3.382, 3.536 and 3.538 
(Bechtold et al. 1995). Strong Mg II and Mg I absorption lines at z = 0.647
have been identified in the quasar spectrum (Angonin - Willaime et al. 1993).

X-ray observations of B1422+231 were made on Jan 14, 1995 for about 21.5ks and 
July 17, 1995 for about 13 ks with the ASCA satellite.
The spectra were extracted from circular regions of 2.5$'$ in radius centered on
B1422+231 and the backgrounds were estimated from similar sized circular regions
located on a source-free region on the second CCD.

We first modeled the spectra of the two observations separately. 
Spectral fits in the observer's frame, incorporating power-law models 
and absorption due to Galactic cold material,
yield photon indices of 1.55$_{-0.08}^{+0.08}$ and 1.46$_{-0.1}^{+0.1}$ for the
Jan 1995 and July 1995 observations respectively. We searched for possible departures
from single power-law models by considering broken power-law models with a 
break energy fixed at 4 keV (rest frame). The Jan 14, 1995 data are 
suggestive of spectral flattening at higher energies while the poor S/N of 
the July 1995 spectrum cannot significantly constrain the spectral slopes.
  
The 2-10keV X-ray fluxes for the Jan 14, 1995 and July 17, 1995 
observations of B1422+231 are estimated to be 1.70$_{-0.37}^{+0.46}$ and 
1.93$_{-0.35}^{+0.40}$ $\times$ 10$^{-12}$ erg s$^{-1}$ cm$^{-2}$ respectively (fits 1 and 4 in Table 5). 

Spectral fits in the quasar mid and high rest-frame bands for the Jan 14, 1995 observation
with  absorbed power-law models and assuming Galactic absorption of 
2.52 $\times$ 10$^{20}$ cm$^{-2}$ yielded spectral indices
of $2.02^{+0.46}_{-0.53}$ and $1.66^{+0.13}_{-0.12}$ respectively.

\subsection{HE1104-1805}
HE1104-1805 is a GL radio-quiet high redshift (z=2.316) quasar with 
an intervening damped Ly$_{\alpha}$ system and a metal absorption system 
at z = 1.66 and a Mg II absorption system at z =1.32. Recent deep near IR
imaging of HE1104-1805 (Courbin, Lidman, \& Magain, 1998) detect
the lensing galaxy at a redshift of 1.66 thus confirming the lens nature of this system.

HE1104-1805 was observed with the ROSAT satellite on June 15 1993 for 13100 sec
and with the ASCA satellite on May 31 1996 for 35989 sec with SIS0
and 35597 sec with SIS1. Reimers et al. (1995) have fit the ROSAT spectrum
of HE1104-1805 in the 0.2 - 2 keV range with an absorbed power law model
and find a photon index of 2.24 $\pm$ 0.16, consistent with our fitted value
of 2.05 $\pm$ 0.2. The main difference between the Reimers et al. and
Chartas 1999 models, used for the fits to the ROSAT spectrum of HE1104-1805,
is that the former model allows the column density to be a free parameter in the
spectral fit while in the latter model the column density is frozen
to the Galactic value of 0.045 $\times$ 10$^{22}$ cm$^{-2}$.
For the data reduction of the ASCA SIS0 and SIS1 observations we extracted
grade 0234 events within circular regions centered on HE1104-1805 and 
with radii of 3.2$'$. The background was estimated by extracting events 
within circular regions in source free areas. 
High, medium and low bit rate data were combined and only Bright mode 
data were used in the analysis. We performed several spectral fits
to the extracted ASCA spectrum with results summarized in Table 6.
A simple spectral fit in the observers frame with an absorbed power-law 
model yields an acceptable fit with a photon index of 1.91$_{-0.06}^{+0.06}$
and a 2-10 keV flux of about 9.4$_{-1.4}^{+1.5}$ $\times$ 10$^{-13}$ erg s$^{-1}$ cm$^{-2}$.
In Figure 3 we show the fit of this model to the ASCA data. 


Spectral fits to the ASCA and ROSAT X-ray spectra 
with absorbed power-law models in the mid and high energy bands 
result in spectral indices of $1.93^{+0.27}_{-0.28}$ and 
$2.01^{+0.1}_{-0.1}$ respectively. For the spectral model we assumed
Galactic absorption with N$_{H}$ = 4.47 $\times$ 10$^{20}$ cm$^{-2}$.

\subsection{The Variable GL BAL Quasar PG1115+080}
Recent observations of PG1115+080 in the FAR - UV with IUE (Michalitsianos et al. 1996)
suggest the presence of a variable BAL region. 
In particular OVI${\lambda}$ 1033 emission and
BAL absorption with peak outflow velocities of $\sim$ 6,000 km s$^{-1}$ 
were observed to vary over timescales of weeks down to about 1 day.
Variations in the BAL absorption features may be due to
changes in the ionization state of the BAL material that could lead to changes
in the column density. A model proposed by Barlow et al. (1992)
to explain the 1 day fluctuations considers the propagation of
an ionization front in the BAL flow.
We expect variations in the BAL column densities to also
manifest themselves as large variations in the observed X-ray flux.
We searched the HEASARC archives and found that PG1115+080 was observed
with the Einstein IPC on Dec 5 1979, the ROSAT PSPC on Nov 21, 1991 and with the ROSAT
HRI on May 27 1994. 
Using the XIMAGE tool {\it detect} on the ROSAT HRI and PSPC images 
of the PG1115+080 observations and searching the NED database we found 
several X-ray sources within a 15 arcmin radius of PG1115+080. Most of these 
sources were detected in the ROSAT International X-ray Optical Survey (RIXOS). 
A list of their coordinates and NED identifications is shown in Table 7.  
The source extraction regions used in the analysis of the PG1115+080 
event files were circles centered on PG1115+080 with radii 
of 1.5 arcmin and 4 arcmin for the PSPC and Einstein observations of
PG1115+080, respectively. We excluded regions containing the nearby RIXOS sources.
The background regions were circles in the near vicinity of PG1115+080. 
We performed various spectral fits to the PG1115+080 data with 
results summarized in Table 8. The observed Einstein IPC and ROSAT PSPC 
spectra of PG1115+080 accompanied by best fit models are shown in Figure 5.
The X-ray observations of PG1115+080 with the ROSAT HRI
show that all the detected X-ray emission is localized within a few
arcsecs. We therefore do not expect any contamination from possible extended lenses.
The HRI image of the field near PG1115+080 is shown in Figure 9.

We modeled the observed spectra as power-laws with Galactic and 
intrinsic absorption. Our spectral fits to the Nov 21, 1991 observation 
imply absorption in excess to Galactic with a modeled
intrinsic absorption of 1.43$_{-1.3}^{+1.3}$ $\times$ 10$^{22}$ cm$^{-2}$ assuming 
a power-law photon index of 2.3 appropriate for high redshift radio-quiet quasars
(Fiore et al. 1998; Yuan et al. 1998). For photon indices 
ranging between 2 and 2.6 the best fit values for the intrinsic absorption
ranges between 0.2 and 3.5 $\times$ 10$^{22}$ cm$^{-2}$.  

To evaluate the statistical significance of the existence 
of intrinsic absorption we calculated the F statistic formed by taking 
the ratio of the difference of $\chi^{2}$ between a fit with only 
Galactic absorption (fit 3 in Table 8) and a new fit that in addition 
to Galactic assumes intrinsic absorption (fit 4 in Table 8)
to the reduced $\chi^{2}$ of the new fit. 
We find an F value of 21 between fits 3 and 4 (see Table 8) implying that the 
addition of an intrinsic absorption component improves the fit to
the Nov 21 1991 observation of PG1115+080 with a probability of exceeding F by chance of about 0.005. 
Our spectral fits to the Dec 5 1979 observations do not indicate absorption in 
excess to the Galactic one. In contrast to the Nov 21 1991 
observation of PG1115+080, the inclusion of intrinsic 
absorption into our model for spectral fits to the Dec 5 1979 
observation produces a significantly larger reduced $\chi^2$.  
The best fit value for the intrinsic absorber column for the Nov 21 1991 observation (fit 2, Table 8)
is poorly constrained to be 1.2$^{+1.1}_{-1.1}$ $\times$ 10$^{23}$ cm$^{-2}$. Notice however from Table 8 that 
this is a very model dependent result.

Our spectral model fits to the presently available X-ray observations of PG1115+080
indicate a decrease of about a factor of 13 of the 0.2-2keV flux between Dec 5 1979
and Nov 21 1991 and an increase by a factor of about 5 between the 
Nov 21 1991 and May 27 1994 observations. Figure 6 shows the estimated 0.2-2keV flux levels 
of PG1115+080 for the three X-ray observations.

The poor S/N of the available spectra make it difficult to discern the cause of the 
X-ray flux variability. Possible origins may include a change in the column density
of the BAL absorber, intrinsic variability of the quasar or a 
combination of both these effects.




\subsection{Q1208+1011, Q1413+117, QJ0240-343}
Q1208+1011 was observed with the ROSAT PSPC on Dec 16 1991 and June 3 1992,
for 2,786 sec and 2,999 sec respectively.
These short observations provide a weak constraint of 2.66$^{+2.1}_{-0.91}$
on the mid band 1-4 keV rest-frame photon index. 
The magnification factor of this lens system is estimated to be approximately 4, 
assuming a singular isothermal sphere lens potential and a lens redshift of z=1.1349 
(Siemiginowska et al. 1998). A recent application of the proximity effect, however,
measured in the Lyman absorption spectrum of Q1208+1011 (Giallongo et al. 1998)
implies an amplification factor as large as 22.   

Q1413+117 is a BAL GL quasar observed with the ROSAT PSPC on July 20, 1991
for 27,863sec. Using the standard detect and spectral fitting software tools
XIMAGE-SOSTA, XIMAGE-detect and XSELECT-XSPEC we detect Q1413+117 at the
3$\sigma$ level in the ROSAT PSPC observation.
The ROSAT PSPC and optical (HST) source coordinates of Q1413+117 are
14h 15m 46.4s, +11$^{\circ}$ 29$'$ 56.3$''$ (J2000),
and 14h 15m 45s, +11$^{\circ}$ 29$'$ 42$''$ (J2000) respectively.
The ROSAT and HST positions are well within the uncertainty of the
ROSAT PSPC pointing accuracy.
The improvements made in the processing of the ROSAT raw data
by the U.S. ROSAT Science Data Center from
the revision 0 product (rp700122) to the revision 2 product (rp700122n00),
used in this analysis, may explain the non-detection of Q1413+117 in the Green \& Mathur 1995 paper.

We fitted the poor S/N PSPC spectrum of Q1413+117 with
a power-law model that included Galactic and intrinsic absorption
due to cold gas at solar abundances. For photon indices ranging between 2.0 
and 2.6 our spectral fits imply intrinsic column densities ranging 
between 2 and 14 $\times$ 10$^{22}$ cm$^{-2}$.     

Recently a pair of bright UV-excess objects, QJ0240-343 A and B, with a separation of 6.1 $''$
were discovered by Tinney (1997). The redshift of both objects was found to be
1.4, while no lens has been detected.
Monitoring of this system in the optical indicates that it
is variable on timescales of a few years. Spectra taken with the 3.9m
Anglo-Australian telescope show a metal-line absorption system at z = 0.543
and a possible system at z = 0.337. QJ0240-343 was observed with the ROSAT PSPC
in January 1992 with a detected count rate of 2.9$\pm$0.7 $\times$ 10$^{-3}$ cnts s$^{-1}$. 
GL theory predicts that the lens for this system lies at about z = 0.5. The geometry of
this system is very similar to that of the double lens Q0957+561.
The large angular image separation of the proposed GL system QJ0240-343
suggests the presence of a lens consisting of a galaxy cluster.
The lens however has yet to be detected and it has been suggested that
this may be a binary quasar system.

\section{DISCUSSION}

\subsection{X-ray Properties of Faint Quasars}

Our present sample of moderate to high S/N ASCA and ROSAT X-ray spectra of GL quasars
contains two radio-loud quasars, Q0957+561 and B1422+231 (see Figures 1 and 2),
and three radio-quiet quasars, HE1104-1805 (see Figure 3), SBS0909+532 (see figure 4) and Q1208+1011.
Derived photon indices in the soft, mid and hard bands for these objects are presented in Table 2.
For the two radio-loud quasars Q0957+561 and B1422+231 we observe a
flattening of the spectra between mid and hard bands while for the 
radio-quiet quasar HE1104-1805 we do not observe any significant change in spectral slope
between mid and hard bands. 
The spectral flattening of radio-loud quasars between mid and hard energy bands 
has been reported for non-lensed quasars (e.g. Wilkes \& Elvis 1987; Fiore et al. 1998; Laor et al. 1997).
The present findings for GL quasars are consistent with those for non-lensed quasars
and imply that the underlying mechanism responsible for the spectral hardening 
in the hard band persists for the relatively high redshift GL quasars of 
our sample with X-ray luminousities that are less 
(by magnification factors ranging between 2 and 30) 
than previously observed objects at similar redshifts.

\begin{figure*}[t]
\plotfiddle{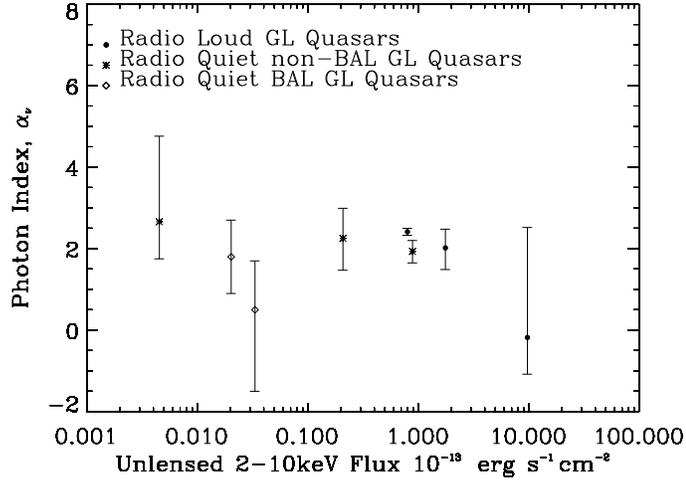}{3.3in}{90}{40.}{40.}{160}{-20}
\protect\caption
{\footnotesize
Photons indices for radio-loud and radio-quiet
GL quasars of our sample as a function of the unlensed 2-10keV flux.
 \label{fig:fig7} 
}
\end{figure*}

\begin{figure*}[t]
\plotfiddle{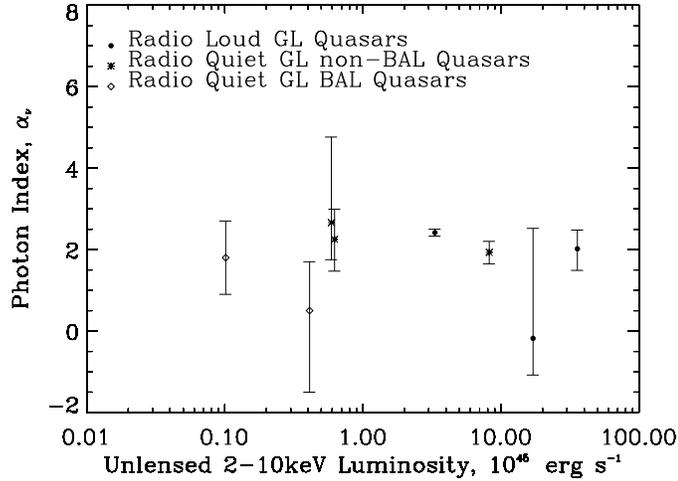}{3.3in}{90}{40.}{40.}{160}{-20}
\protect\caption
{\footnotesize 
Photons indices for radio-loud and radio-quiet
GL quasars of our sample as a function of the 2-10keV 
luminosity. X-ray luminosities have been corrected for the magnification 
effect.
 \label{fig:fig8} 
}
\end{figure*}

Our analysis makes use of the GL amplification effect
to extend the study of quasar properties to X-ray flux levels
as low as a few $\times$ 10$^{-16}$ erg s$^{-1}$ cm$^{-2}$. The limiting sensitivity of
the ROSAT All-sky Survey, for example, on which many recent studies are based,
is a few $\times$ 10$^{-13}$ erg s$^{-1}$ cm$^{-2}$.
We find that the spectral slopes of the radio-loud and non BAL quiet-quasars
of our sample are consistent with those found in quasars of higher flux levels 
and do not appear to approach the observed spectral index of $\sim$ 1.4 
of the hard X-ray background. 
Absorption due to known intervening systems in Q0957+561, B1422+231,
SBS0909+532, Q1208+1011 and HE1104-1805 apparently does not lead to the spectral hardening observed in
the Vikhlinin et al. (1995) sample at flux levels below 
$\sim$ 10$^{-13}$ erg s$^{-1}$ cm$^{-2}$.  
However, we do find that modeling the two radio-quiet BAL quasars PG1115+080 and Q1413+117
and the radio-loud quasar PKS1830-211, which shows strong X-ray absorption
(Mathur et al. 1997), with simple power-law models with Galactic absorption
results in very low spectral indices (see Table 2).
Similar unlensed sources will therefore contribute to the remaining unresolved 
portion of the XRB.
The presently available sample size of X-ray detected BAL radio-quiet quasars, however,
will have to be significantly increased before we can make a statisticly significant 
quantitative assessment of the BAL quasar contribution to the hard XRB.

\subsection{X-ray Properties of Gravitationally Lensed BAL Quasars}
Approximately 10\% of optically selected quasars have optical/UV
spectra that show deep, high-velocity Broad Absorption Lines (BAL)
due mostly to highly ionized species such as C IV, Si IV, N V and O VI.
However, a small fraction of BAL quasars show low ionization transitions of
Mg II, Al III, Fe II and Fe III as well (e.g. Wampler et al. 1995).
The observed absorption troughs are found bluewards of the associated resonance
lines and are attributed (see Turnshek et al. 1988)
to highly ionized gas flowing away from the central source at speeds ranging
between 5,000 and 30,000 km s$^{-1}$.
Recent polarization observations (Goodrich 1997)
indicate that the true fraction of BAL's and BAL covering
factors may be substantially larger ($>$ 30\%) than the presently quoted value
of 10\%. Only a very small number of BAL quasars and AGN have been reported in the literature
with detections in the X-ray band (PHL5200, Mrk231, SBS1542+541, 1246-057, and Q1120+0195(UM425)).
With this work we also add to the list of X-ray detected BAL quasars the GL quasars 
PG1115+080, Q1413+117  and possibly RXJ0911.4+0551.  We consider PG1115+080 and RXJ0911.4+0551 
intermediate BAL quasars because of the relatively low peak velocities of the ouflowing absorbers
(see Table 9 for a list of outflowing velocities of GL BAL quasars).
The X-ray spectra obtained from X-ray observations of BAL quasars
have modest (PHL5200, \& Mrk231) to poor S/N and cannot accurately constrain the
BAL column densities.

\begin{figure*}[t]
\plotfiddle{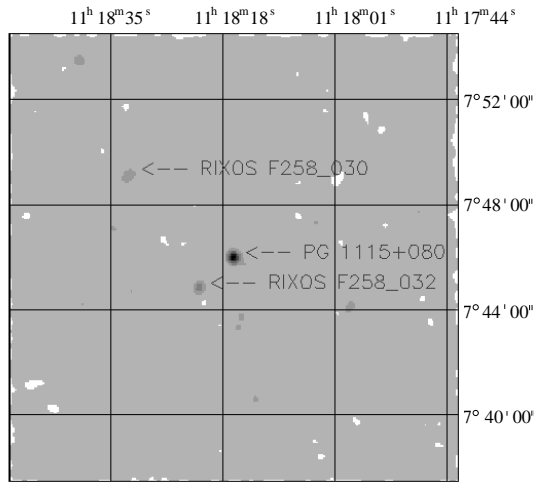}{3.2in}{-90}{43.}{43.}{-160}{240}
\protect\caption
{\footnotesize 
HRI image of the field near PG1115+080.
 \label{fig:fig9} 
}
\end{figure*}

\begin{figure*}[t]
\plotfiddle{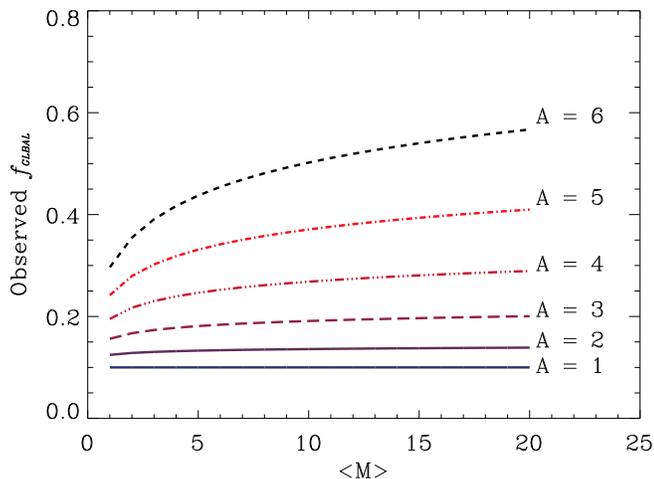}{3.2in}{90}{40.}{40.}{160}{-20}
\protect\caption
{\footnotesize
Plot of the estimated observed fraction of
GL BAL quasars as a function of attenuation value A and
magnification factor $<M>$.
 \label{fig:fig10}
}
\end{figure*}

Several of the GL quasars of our sample are known to contain intervening and intrinsic
absorption. 
In particular PG1115+080 is known to contain a variable BAL system (Michalitsianos et al. 1996). 
The available X-ray observations indicate that PG1115+080 is a highly variable X-ray 
source.
The large X-ray flux variations (a factor of about 13 decrease in X-ray flux  between 
December 5 1979 and November 21 1991 and about a factor of 5 increase between
November 21 1991 and May 27 1994) may possibly be used to substantially reduce the errors
in  the determination of the time delay of this GL system. 
Such a monitoring program will have to await the launch of the Chandra 
X-ray Observatory (CXO), a.k.a. AXAF, which has the spatial resolution to resolve the 
lensed images. 

The GL quasars Q1413+117, Q1120+0195 and RXJ0911.4+0551
have also been detected in X-rays and are known to contain BAL features. Unfortunately  
the available X-ray data have poor S/N and we have only provided estimates of their 
X-ray flux and luminosity. Recent high resolution optical and NIR imaging of
RXJ0911.4+0551 have resolved the object into
four lensed images and a lensing galaxy (Burud et al. 1998).
They also detect a candidate galaxy cluster 38$''$ away from
the image A1 with an estimated redshift of 0.7. 
It is possible that a large fraction of the detected X-ray emission 
in the ROSAT HRI observations of RXJ0911.4+0551
is originating from the cluster of galaxies. 

An interesting finding made by searching through the literature is the 
apparantly large fraction of optically detected GL BAL quasars. In particular we 
found seven GL BAL quasars out of a total of about 20 radio-quiet GL quasar candidates
known to date. The probability of finding 7 or more BAL
quasars out of a sample of 20 GL radio-quiet quasars assuming a true BAL fraction
(amongst radio-quiet quasars only) of 0.11 is about 4 $\times$ 10$^{-3}$.
In Table 9 we list several properties of these GL BAL quasars. 
Thus we find that at least 35$\%$ of radio-quiet gravitationally lensed quasars
contain BAL features which is significantly larger than the 10$\%$ fraction of BAL quasars found
in optically selected quasar samples (almost all BAL's are radio-quiet and about
90$\%$ of optically selected quasars are radio-quiet).
Recently, BAL's have also been identified in a few radio-loud quasars 
(Brotherton et al. 1998).
These observations suggest that a large fraction of BAL
quasars are missed from flux limited optical surveys,
a view that has also been proposed by Goodrich (1997)
based on polarization measurements of BAL quasars.
One plausible explanation for the over-abundance of BAL quasars amongst radio-quiet
GL quasars is based on the GL magnification effect which causes the 
luminosity distributions of BAL quasars and GL BAL quasars to 
differ considerably such that
presently available flux limited surveys of BAL quasars
detect more GL BAL quasars.
We have created a simple model that can explain the difference
between the observed GL BAL fraction of $\sim$ 35\% and
the observed non-lensed BAL quasar fraction of $\sim$ 10\%.\\
Our model makes use of the quasar luminosity function as parameterized by Pei (1995),
assumes that only 20\% of BALs observed in optical surveys
of unlensed quasars are attenuated by a factor A (see Goodrich, 1997) and it 
uses the Warren et al. (1994) optical limits
for non-lensed quasars and the CASTLES survey optical limits
for lensed quasars (Kochanek et al. 1998).
To simplify the analysis we assume an average magnification factor of $<$M$>$
for the GL quasars rather than incorporate each magnification factor separately.

A  survey of lensed quasars, with luminosity limits between L$_{1}$ and L$_{2}$,
will have a {\it true} luminosity range, assuming an average lens magnification factor
of $<$M$>$, that lies between ${L_{1}}\over{<M>}$ and ${L_{2}}\over{<M>}$ for unattenuated lensed BAL quasars
and that lies between ${L_{1}A\over<M>}$ and ${{L_{2}A}\over{<M>}}$ for attenuated lensed BAL quasars.
Following the arguments of Goodrich (1997) we assume that only an
observed fraction of 20\% of BAL quasars are attenuated (this is 
approximately the observed fraction of  BAL quasars with significant polarization).
Based on what we have just discussed, the observed fraction , $f_{ogb}(L_{1},L_{2})$, 
of GL BAL quasars in the luminosity 
range of L$_{1}$ to L$_{2}$ can be approximated
with the observed fraction, $f_{ob}({{L_{1}}\over{<M>}},{{L_{2}}\over{<M>}})$, of non-GL 
BAL quasars in the {\it observed} luminosity range of ${{L_{1}}\over{<M>}}$ to ${{L_{2}}\over{<M>}}$.

\begin{equation}
f_{ogb}(L_{1},L_{2}) = f_{ob}({{L_{1}}\over{<M>}},{{L_{2}}\over{<M>}})
\end{equation}

\noindent
We separate the observed BAL quasar fraction, $f_{ob}({{L_{1}}\over{<M>}},{{L_{2}}\over{<M>}})$, 
into the fraction that is attenuated, $f_{oba}({{L_{1}}\over{<M>}},{{L_{2}}\over{<M>}})$, and the fraction
$f_{obna}({{L_{1}}\over{<M>}},{{L_{2}}\over{<M>}})$ that is not attenuated,

\begin{eqnarray}
f_{ob}({{L_{1}}\over{<M>}},{{L_{2}}\over{<M>}}) =  \nonumber \\
f_{oba}({{L_{1}}\over{<M>}},{{L_{2}}\over{<M>}}) + \nonumber \\
f_{obna}({{L_{1}}\over{<M>}},{{L_{2}}\over{<M>}})
\end{eqnarray}

\noindent
If we assume that the luminosity distribution of non-attenuated BAL quasars 
is similar to that of non BAL quasars we expect the fraction 
$f_{obna}({{L_{1}}\over{<M>}},{{L_{2}}\over{<M>}})$ to be independent of
luminosity range and therefore approximately equal to the observed value 
$f_{obna}({L_{3}},{L_{4}})$ $\sim$ 8$\%$, where L$_{3}$ and L$_{4}$
are the Warren et at. (1994) optical luminosity limits.

As pointed out by Goodrich 1997 the attenuation 
expected to be present in about 20$\%$ of all BAL
quasars causes the observed luminosity function for BAL quasars to be 
considerably different from the true luminosity function for BAL
quasars.

The ratio of observed, $f_{oba}$, to true, $f_{tba}$, fraction of attenuated 
BAL quasars can be determined if one incorporates the effect
of attenuation in the quasar luminosity function.
In particular if we define $N({{L_{1}}\over{<M>}},{{L_{2}}\over{<M>}}z_{1},z_{2})$ as the integral of the quasar luminosity
function as parametrized by Pei (1995) over the luminosity 
range ${L_{1}}\over{<M>}$ and ${L_{2}}\over{<M>}$ and the redshift range
$z_{1}$ and $z_{2}$ then we may write the ratio of observed to true fraction of attenuated
BAL quasars within this luminosity range as,

\begin{equation}
{{f_{oba}( {{L_{1}}\over{<M>}},{{L_{2}}\over{<M>}} ) }\over{f_{tba}}}=
{{N({{L_{1}A}\over{<M>}},{{L_{2}A}\over{<M>}},z_{1},z_{2})}\over{N({{L_{1}}\over{<M>}},{{L_{2}}\over{<M>}},z_{1},z_{2})}}
\end{equation}

The observed fraction of about 2\% of attenuated non-lensed BAL quasars, 
however, is measured within the Warren et al. (1994) optical limits of
L$_{3}$ = 2.7 $\times$ 10$^{46}$ erg s$^{-1}$ and L$_{4}$ = 3.8 $\times$ 10$^{47}$ erg s$^{-1}$
and redshift range of z$_{3}$ = 2 and z$_{4}$ = 4.5.
We therefore write the ratio of observed to true fraction of attenuated
BAL quasars within the L$_{3}$ and L$_{4}$ range as,

\begin{equation}
{{f_{oba}( {{L_{3}}},{{L_{4}}} ) }\over{f_{tba}}}=
{{N({{L_{3}A}},{{L_{4}A}},z_{3},z_{4})}\over{N({{L_{3}}},{{L_{4}}},z_{3},z_{4})}}
\end{equation}

Combining equations 4, 5, 6 and 7 we obtain the following  expression for the observed 
fraction of GL BAL quasars as a function of average GL magnification $<M>$
and BAL attenuation factor A,

\begin{eqnarray}
f_{ogb}(L_{1},L_{2})= f_{obna}(L_{3},L_{4}) + {f_{oba}(L_{3},L_{4})} \nonumber \\ 
{\times}{{N({{L_{3}}},{{L_{4}}},z_{3},z_{4})}\over{N({{L_{3}A}},{{L_{4}A}},z_{3},z_{4})}}
{{N({{L_{1}A}\over{<M>}},{{L_{2}A}\over{<M>}},z_{1},z_{2})}\over{N({{L_{1}}\over{<M>}},{{L_{2}}\over{<M>}},z_{1},z_{2})}}
\end{eqnarray}

In Figure 10 we plot the expected observed fraction of
GL BAL quasars as a function of attenuation values A and 
magnification factors $<M>$.

The magnification effect of GL quasars alone cannot explain the observed 
enhanced GL BAL quasar fraction of $\sim$ 35$\%$. 
By combining, however, the magnification effect with the 
presence of an attenuation of the continuum in a fraction of 
BAL quasars, as suggested by the polarization observations by 
Goodrich 1997, our simple model can reproduce the observed GL BAL
quasar fraction of $\sim$ 35$\%$. For a range of average magnification factors
$<M>$ between 5 and 15 we obtain attenuation values A ranging between 5
and 4.5. The range of attenuation values of 4.5 to 5, suggested by the 
observed fraction of GL BALQSO's, is close to the range of 3 to 4 
implied by the observed polarization distributions
of BALQSO's and non-BAL radio-quiet quasars (Goodrich 1997), especially
considering the uncertainties in both analyses.
A value of $<M>$ $\sim$ 10 is consistent with typical estimated 
values for GL quasars, (see, for example, our GL model estimates in Table 3).


\section{CONCLUSIONS}

We have introduced a new approach in studying the X-ray properties 
of faint quasars. Our analysis makes use of the GL amplification effect
to extend the study of quasar properties to X-ray flux levels
as low as a few $\times$ 10$^{-16}$ erg s$^{-1}$ cm$^{-2}$.
For the two radio-loud GL quasars Q0957+561 and B1422+231 we observe a
flattening of the spectra between mid and hard bands (rest-frame)  while for the radio-quiet
quasar HE1104-1805 we do not observe any significant change in spectral slope
between mid and hard bands.
The present findings in GL quasars are consistent with those of non-lensed quasars
and imply that the underlying mechanism responsible for the spectral hardening
from mid to hard bands persists for the relatively high redshift GL radio-loud quasars 
of our sample with X-ray luminousities that are less (by the magnification factors indicated in Table 3) 
than previously observed objects at similar redshifts.
 
Our results suggest that radio-loud and non-BAL radio-quiet quasars with unlensed fluxes
as low as a few $\times$ 10$^{-16}$ erg s$^{-1}$ cm$^{-2}$  
do not have spectral slopes that are any different from brighter quasars.
Modeling the spectra of the two GL BAL quasars and the radio-loud quasar PKS1830-211 in our sample
with simple power-laws and including only Galactic absorption leads to spectral
indices that are considerably flatter than the average values for quasars. 
These results therefore imply that BAL quasars and quasars with associated absorption will contribute
to the unresolved portion of the hard XRB.
We must emphasize, however, that our present sample of GL quasars will need to be enlarged to
assess the significance of the contribution of BAL quasars to the XRB. 
X-ray observations in the near future with the X-ray missions CXO, XMM and ASTRO-E
will significantly aid in adding many more GL quasars to this sample.

Our analysis of several X-ray observations of the GL BAL quasar PG1115+080   
show that it is an extremely variable source. Fits of various models
to the spectra obtained during these observations suggest that
the X-ray variability is partly due to a variable BAL absorber. 
The X-ray flux variability in this source can be used to
improve present measurements of the time delay.
The large variability in the X-ray compared to optical band offers the prospect 
of substantially reducing the errors in deriving a 
time delay from cross-correlating image light curves.   
A precise measurement of the time delay combined with an accurate model for the mass 
distribution of the lens can be used to derive a Hubble constant that does not depend 
on the reliability of a ``standard candle''. 
The scheduled monitoring of PG1115+080 with the CXO
will provide spatially resolved spectra and light curves for the individual lensed images.

One of the significant findings of this work was a surprisingly 
large fraction of BAL quasars that are gravitationally lensed.
In particular we find 7 BAL quasars out of a sample of 20 GL radio-quiet quasars.
We have successfully modeled this effect and find 
that an attenuation factor A $\sim$ 5  of the BAL continuum of
only 20$\%$ of all BAL quasars is consistent with the observed GL fraction of 35$\%$.
We emphasize that the magnification effect alone cannot explain
the observed difference between BAL fractions for lensed
and non-lensed quasars. One needs to incorporate in addition
an attenuation mechanism to produce the observed results.
These observations therefore are suggestive of the existence
of a hidden population of absorbed high redshift quasars which have eluded
detection by present flux limited surveys. 
As X-ray and optical surveys approach lower flux limits we expect 
the fraction of BAL quasars found to increase.

I would like to thank N. Brandt, M. Eracleous, G. Garmire, and J. Nousek for helpful discussions and comments.
This work was supported by NASA grant NAS 8-38252. 
This research has made use of data obtained through the 
High Energy Astrophysics Science Archive Research Center 
Online Service, provided by the
NASA/Goddard Space Flight Center.


\end{document}